\documentclass[a4paper]{jpconf}
\usepackage{graphicx}
\usepackage{amssymb}
\usepackage[intlimits]{amsmath}
\usepackage{amsmath}
\begin{document}

\title{Investigation of Nonlinear Communication Channel with Small Dispersion via Stochastic Correlator Approach}

\author{A V Reznichenko$^{1,2}$, I S Terekhov$^{1,2}$}
\address{$^1$ Theory department, Budker Institute of  Nuclear
Physics of Siberian Branch Russian  Academy of Sciences, Novosibirsk 630090, Russia
}
\address{$^2$  Novosibirsk State University, Novosibirsk 630090, Russia}

\ead{a.v.reznichenko@inp.nsk.su}

\begin{abstract}
We consider the optical fiber channel modelled by the nonlinear Schr\"{o}dinger
equation  with  additive white Gaussian noise and with large signal-to-noise ratio. For the small dispersion case we present the approach to analyze the stochastic nonlinear Schr\"{o}dinger equation. Taking into account the averaging procedure (frequency filtering) of the output signal detector we find the first  corrections in small dispersion parameter to the correlators of the input signal recovered by the backward propagation. These correlators are the important ingredients for the calculation of the channel capacity and the optimal input signal distribution. We assert that the information channel characteristics essentially depend on the procedures of the output signal filtering and the recovery of the transmitted signal.

\end{abstract}

%===================================
\section{Introduction}

The problem of analytical calculations of characteristics for a given information channel is of great importance in the information theory due to the practically significant applications of optical fibers. These characteristics are as follows: the conditional probability density function ($P[Y|X]$), i.e., the probability to detect the output signal $Y(t)$ if the transmitted input signal is $X(t)$; the distribution of the output signal ($P_{out}[Y]$); the distribution of the recovered input signal, the output ($H[Y]$) and conditional ($H[Y|X]$) entropies, the mutual information ($I_{P_X}=H[Y]-H[Y|X]$); the optimal input signal distribution; the channel capacity ($C = \max_{P_{X}[X]} I_{P_{X}[X]}$) and others. In our work we consider the problem of the signal propagation in a noisy information channel where the propagation is governed by the stochastic nonlinear Schr\"{o}dinger equation (NLSE) with the additive white Gaussian noise for the case of small dispersion and the large  signal-to-noise power ratio ($\mathrm{SNR}$).

%===============Model
We consider the complex signal $\psi(z,t)$ which is related with the electric field in the optical fiber as $\vec{E}(z,t) \propto \vec{e}_0 \mathrm{Re}\Big\{\psi(z,t+\frac{z}{v_g}) e^{i k_0 z -i \omega_0 t} \Big\}$ with  $\omega_0=\omega(k_0)$ and $v_g=\frac{d \omega}{d k}(k_0)$  being the carrier frequency and the group velocity, correspondingly. The propagation of the signal $\psi(z,t)$  is described by the NLSE
with the additive white Gaussian noise, see {\cite{Haus:1991,Mecozzi:1994,Iannoe:1998,Turitsyn:2000}}:
\begin{eqnarray}\label{startingCannelEqt}
&&\!\!\!\partial_z \psi(z,t)+i\beta\partial^2_{t}\psi(z,t)-i\gamma |\psi(z,t)|^2 \psi(z,t)=\eta(z,t) \,,
\end{eqnarray}
here $\beta$ is the second dispersion coefficient, $\gamma$ is the Kerr nonlinearity
coefficient, $\eta(z,t)$ is the additive complex white noise with zero mean $\langle
\eta (z,t)\rangle_{\eta}=0$ and correlation function
\begin{eqnarray}\label{noisecorrelatort}
\langle \eta (z,t)\bar{\eta}(z^\prime,t^\prime)\rangle_{\eta} =  Q
\delta(z-z^\prime)\delta(t-t^\prime)\,,
\end{eqnarray}
where bar means complex conjugation, $Q$ is a power of the white Gaussian noise
per unit length and per unit frequency interval. The input condition for the
signal $\psi(z,t)$ is $\psi(z=0,t)=X(t)$ and the output one is $\psi(z=L,t)=Y(t)$, where $Y(t)$ is determined both the input condition and the noise impact in Eq.~(\ref{startingCannelEqt}). The frequency bandwidth $W'$ of the noise $\eta(z,t)$ is assumed to be much greater than the frequency bandwidth $W$ of the input signal $X(t)$: $W' \gg W$.

There are different approaches to analyze
Eq.~(\ref{startingCannelEqt}) and to find informational channel
characteristics, primarly, the conditional probability density function
$P[Y|X]$ that is one of the most important characteristics. The field theory
approach is based on the path-integral formulation of the quantity
$P[Y|X]$ \cite{MSR:1973, Terekhov:2014}. We have performed some
estimations for the channel with the small dispersion parameter via
this approach \cite{Reznic:2017IEEE}.  Now we proceed the small
dispersion analysis of the Eq.~(\ref{startingCannelEqt}) by
exploiting the stochastic approximation of the noisy equation
~(\ref{startingCannelEqt}) within the detector averaging procedure.
Firstly, we perform the linearization of the
Eq.~(\ref{startingCannelEqt}) using the fact that for the large
$\mathrm{SNR}$ case the solution of Eq.~(\ref{startingCannelEqt})
can not significantly deviate from the solution (we denote this solution as $ \Phi(z,t)$) of  this equation
with zero noise  and with the input
condition $\Phi(z=0,t)=X(t)$. Thus we consider the linear equation
for the difference $F(z,t) \propto \psi(z,t)-\Phi(z,t)$, see
Eq.~(\ref{Fequation}) below in the Sec. \ref{Secoutputfilt}.
Secondly, the averaging action  of the output signal detector allows
us to reduce this linear equation to one that can be treated with
the perturbation theory both in small noise parameter
($1/\mathrm{SNR} \ll 1$) and small dispersion parameter
$\widetilde{\beta}=\beta L ({W}/{(2\pi)})^2 \ll 1$: see
Eq.~(\ref{Faequation}) below in the Sec. \ref{Secoutputfilt}.  The
detector at $z=L$ performs the frequency filtering with bandwidth
$W_{a}$ which obeys the inequality $W \ll W_a \ll W'$. Therefore we
have the separation of the time scales for the input signal function
$X(t)$ (``slowly varying function''), averaged function $F(z,t)$,
and the noise $\eta(z,t)$. This separation enables us to develop the
perturbation theory in small dispersion parameter. We demonstrate
how the channel informational characteristics depend on the procedure
of the output signal detecting (frequency filtering) and the
algorithm of the input signal recovery on the base of the filtered
output signal, see Sec. \ref{SecInpusignal}. Finally, we present the  correlators of the input signal
$\widetilde{X}(t)$ recovered from the noisy channel and the conditional probability density function up to the first corrections in dispersion parameter, see Sec. \ref{SecPCC}.

%===================================
\section{Input signal model}
\label{SecInput signal}

We assume the following simple coding model for the input signal $X(t)$:
\begin{eqnarray}\label{Xtmodelg}
X(t)=\sum^{N}_{k=-N} C_{k} \, g_{k}(t), \qquad g_{k}(t)=g_{0}(t-k T_0),
\end{eqnarray}
where the whole (large) time interval $T=(2N+1)T_0$ is divided into $(2N+1)$ subintervals of the  duration $T_0$, and  we have (almost) nonoverlapping envelope functions $g_{k}(t)=g_{0}(t-k T_0)$ with the bandwidth of order of $1/\widetilde{\delta t}=W/(2\pi)$ in every subinterval $[-T_0/2+k T_0, T_0/2+k T_0]$. For simplicity we choose Gaussian  functions as the envelope shape:
\begin{eqnarray}\label{Xtmodel}
g_{0}(t)= \sqrt{\frac{T_0}{\sqrt{\pi}\widetilde{\delta t}}}\exp\left\{-\frac{t^2}{2 \widetilde{\delta t}^2}\right\},
\qquad \hat{g}_{0}(\omega)=\int dt e^{i \omega t}g_{0}(t)=\sqrt{2\sqrt{\pi}\widetilde{\delta t} T_0}\exp\left\{-\frac{ \widetilde{\delta t}^2\omega^2}{2}\right\}.
\end{eqnarray}
If the sparsity is large: $T_0/\widetilde{\delta t}=N_{t} \gg 1$, i.e, the overlapping  is small, and we have $g_{k}(t) g_{k'}(t)\approx \delta_{k,\,k'}g^2_{k}(t)$. For our estimations we choose $N_{t}=5$ (practically, if $N_{t}=5$ the overlapping, i.e. $\int dt g_k(t) g_{k+1}(t)/\int dt g^2_k(t)$, is about $0.002$). The function $g_{0}(t)$ is normalized by the condition $\int^{\infty}_{-{\infty}}{dt} g^2_k(t)/{T_0}=1$. When recovering the input signal the coefficients $C_{k}$ can be found as the projection of the signal $X(t)$ on basis functions:
\begin{eqnarray}\label{Ckrecover}
C_{k}\equiv r_{k}  \exp\{i \phi_k \} \sqrt{P_{ave}}=\int_T \frac{dt}{T_0} g_k(t) X(t).
\end{eqnarray}

In our model we suppose coefficients $C_{k}$ to have various amplitude with given distribution, $C_{k}= r_{k}  \exp\{i \phi_k \} \sqrt{P_{ave}}$, where we separate the dimensionless random real value $r_{k}$ and the phase $\phi_k$. We suppose that the phases $\phi_k$ are constants over the subinterval $T_0$ and assume a discrete values, say, from the set $\{0,\,\pi/2, \pi, 3\pi/2\}$. In the representation $C_{k}=\sqrt{P_{ave}} r_{k}  \exp\{i \phi_k \}$ the dimension parameter $P_{ave}$ is the average power over the given input signal distribution $P[C]$:
\begin{eqnarray}\label{PaveC}
&&\!\!\!\!\!\!P_{ave}=\int \!\!{\cal D}C P[C] \frac{1}{T}\int_{T}
 |X(t)|^2 dt =\int \prod^{N}_{j=-N} \Big(|C_k|d|C_k| d \phi_k \Big)  P[C] \frac{1}{2N+1}\sum_{k=-N}^N |C_k|^2.
\end{eqnarray}
In our calculations we present the input signal in the form $X(t)=\rho(t) \exp\{i \phi_0(t)\}$, where $\rho(t)=|X(t)|\approx \sqrt{P_{ave}} \sum^{N}_{k=-N}  r_k g_{k}(t)$, and the constant over the subintervals $T_0$  phase $\phi_0(t)$ is smoothed function in such a way that all time derivates ($\dot{\phi _{0}}$, $\ddot{\phi _{0}}$) are localized on the borders of subintervals $T_0$, and if one neglects the overlapping then $\dot{\phi _{0}}(t) g_{k}(t)$, $\ddot{\phi _{0}}(t) g_{k}(t)$ may be considered as zero.

We will use the dimensionless parameter  $\mu(t)=\gamma L\, \rho^2(t) $
characterizing the impact of  nonlinearity in the phase evolution for the NLSE
(\ref{startingCannelEqt}) with $\beta=0$, see \cite{TTKhR:2015,Terekhov:2017}. The dimensionless parameter
$\widetilde{\gamma}=\gamma L P_{ave}$ is the average nonlinearity parameter. We use the dimensionless dispersion parameter $\widetilde{\beta}={\beta L}/{\widetilde{\delta t}^2}=\beta L ({W}/{(2\pi)})^2$, where $W/{(2\pi)}$ is referred to as the frequency bandwidth of the input signal $X(t)$ (and $g_{k}(t)$), see Eq.~(\ref{Xtmodel}). In our perturbative considerations the parameter $\widetilde{\beta}$ is assumed to be small: $\widetilde{\beta} \ll 1$.

%===================================
\section{Output signal filtering and stochastic NSLE solution}
\label{Secoutputfilt}

The concrete form of the conditional probability density function
$P[Y|X]$  depends essentially on the signal detection procedure: this
implies the output signal detection and the reconstruction of the
input signal on the base of the detected output signal. Let us first
consider the procedure of the output signal $\psi(z=L,t)$ detection.
%==================
\subsection{Linearization}
We present the NLSE solution $\psi(z,t)$ in the form
\begin{eqnarray}\label{psi}
\psi(z,t)=\Phi(z,t)+e^{i \theta_{0}(z,\,t)} F(z,t),
\end{eqnarray}
where $\Phi(z,t)$  is the solution of the NLSE equation  (\ref{startingCannelEqt}) with zero noise, with nonzero $\beta$, and with the input boundary condition $\Phi(z=0,t)=X(t)=\rho(t) \exp\left\{i \phi_{0}(t)\right\}$. It means that $F(z=0,t)=0$. The phase $\theta_{0}(z,\,t)=\phi_{0}(t)+\mu(t){z}/{L}$ corresponds to the phase of the solution
$\Phi^{(0)}(z,t)$ which is the NLSE solution with zero noise and zero $\beta$ and with the input boundary condition: $\Phi^{(0)}(z=0,t)=\rho(t)e^{i \phi_{0}(t)}$, see \cite{TTKhR:2015}.
For nonzero $\beta$ the  solution  $\Phi(z,t)$ has the following
form in $\beta$ expansion: $\Phi(z,t)=e^{i
\theta_{0}(z,\,t)}\sum^{\infty}_{n=0} \widetilde{\Phi}^{(n)}(z,t)$.
Quantities $\widetilde{\Phi}^{(n)}(z,t)$   are polynomials in nonlinear parameter $\mu(t){z}/{L}$ up to
the factors $(z \beta)^n$.
The function $\widetilde{\Phi}^{(0)}(z,t)=\rho(t)$ corresponds to
the leading order in small $\widetilde{\beta}$, and
$\widetilde{\Phi}^{(1)}(z,t)$ corresponds to the next-to-leading
order, and so on. As an example, the first correction reads
$\widetilde{\Phi}^{(1)}(z,t)=\beta z
\sum^{2}_{k=0}\phi^{(1)}_{k}(t)\left(\mu(t){z}/{L}\right)^k$, where
$\phi^{(1)}_{0}(t)=2 \dot{\rho}\, \dot{\phi_{0}} +\ddot{\phi_{0}}\,
\rho +i \left(\rho \dot{\phi _{0}}^2 -\ddot{\rho} \right)$,
$\phi^{(1)}_{1}(t)=\ddot{\rho}+3 {\dot{\rho}^2}/{\rho}+i\left(
\ddot{\phi_{0}} \, \rho+4 \dot{\rho} \dot{\phi_{0}}  \right)$,
$\phi^{(1)}_{2}(t)=i \frac{2}{3} \left[\ddot{\rho}+5
{\dot{\rho}^2}/{\rho}\right]$. In the $\beta$ expansion of the NLSE
equation (\ref{startingCannelEqt}) with zero noise it is easy to
find $\widetilde{\Phi}^{(n)}(z,t)$ for arbitrary  $n$.

We assume that the function $F(z,t)$ is of order of $\sqrt{Q}$ (i.e., we consider only such  output signals $Y(t)$ that $|Y(t)-\Phi(z=L,t)| \sim 1/\sqrt{\mathrm{SNR}}$) and  therefore $F(z,t) \ll \Phi(z,t)$ in Eq.~(\ref{psi}).  We can obtain from Eq.~
(\ref{startingCannelEqt}) the  linear equation for $F(z,t)$
up to terms ${\cal O}(\beta^2)$ and ${\cal O}(\mathrm{SNR}^{-1})$:
\begin{eqnarray}\label{Fequation}
&&\!\!\!\!\!\!\partial_z F(z,t)+i\beta\partial^2_{t}F(z,t)-i\frac{\mu(t)}{L}\left(F(z,t)+\overline{F}(z,t)\right)\approx\widetilde{\eta}(z,t)-2 i \beta \Big(\partial_{t}\theta_{0}(z,\,t)\Big) \partial_{t}F(z,t)+  \\&& \!\!\!\!\!\!2 i \gamma \rho(t)
\left[ 2F(z,t) Re[\widetilde{\Phi}^{(1)}(z,t)]+\overline{F}(z,t)\widetilde{\Phi}^{(1)}(z,t) \right]-i \beta  F(z,t) \left[i \Big(\partial^2_{t}\theta_{0}(z,\,t)\Big)- \Big(\partial_{t}\theta_{0}(z,\,t)\Big)^2  \right]  \nonumber ,
\end{eqnarray}
where in $\Phi(z,t)$ we have taken into account only the first
correction in ${\beta}$. Here $\widetilde{\eta}(z,t)=e^{-i
\theta_{0}(z,\,t) }{\eta}(z,t)$ is the noise with the same
statistics as in equation (\ref{noisecorrelatort}) and we can
replace $\widetilde{\eta}(z,t)$  with ${\eta}(z,t)$.
%==================
\subsection{Averaging of the detector}
For simplicity we assume that the output signal detector performs
the following averaging:
\begin{eqnarray}\label{OSaveraging}
&&F(z,t) \rightarrow F_{(a)}(z,t)=\int^{t+\tau_a}_{t-\tau_a} \frac{dt'}{2\tau_a}F(z,t'), \qquad  \hat{F}_{(a)}(z,\omega) = \hat{F}(z,\omega)\, \mathrm{sinc}[\omega \tau_a],
\end{eqnarray}
where the averaging time parameter $\tau_a=2\pi/W_a$ is connected with the output frequency filtering bandwidth $W_a$. To generalize this procedure, we can introduce $\hat{F}_{(a)}(z,\omega)=\hat{F}(z,\omega) k_{\tau_a}(\omega)$, where filtering function  $k_{\tau_a}(\omega)$ truncates the output signal up to the given bandwidth $W_a=2\pi/\tau_a$ ($k_{\tau_a}(\omega)$ is localized on the bandwidth $W_a$).

For sufficient small $\widetilde{\beta}$ we can choose  $\tau_a$ in the wide region in such a way that the parameter $\widetilde{\beta}_{a}=\beta L /(\widetilde{\delta t}\tau_a)$ is small:
\begin{eqnarray}\label{taurelation}
\beta L/\widetilde{\delta t}  \ll \tau_a \ll \widetilde{\delta t},
\qquad  \mathrm{SNR}^{-1} \ll \widetilde{\beta}  \ll \widetilde{\beta}_{a} \ll 1, \qquad
\beta L/\tau^2_a  \gtrsim 1.
\end{eqnarray}
These relations (\ref{taurelation}) allows us  to  perform the
averaging of the equation (\ref{Fequation}). From this equation one
can see that $F(z,t)$ has the time scale of order of $\Delta
=2\pi/W'$: the frequency bandwidth of $F(z,t)$ is determined by the
bandwidth of the noise in the right hand side, $W' \gg W$).
The frequency bandwidth of the averaged solution, see
Eq.~(\ref{OSaveraging}), is $W_a$: $W \ll W_a \ll W'$. The last relation in
Eq.~(\ref{taurelation}), i.e., $\beta L/\tau^2_a  \gtrsim  1$, implies
that $\widetilde{\beta}^2_{a} \gtrsim \widetilde{\beta}$.

The functions  $\mu(t)$, $\rho(t)$, $e^{i \phi_0(t)}$ and their time derivatives  can be
considered as  slowly varying functions in comparison with both
$F(z,t)$ and the averaged solution $F_{(a)}(z,t)$: we generally
denote them as $s(t)$. This means that we can average the equation
(\ref{Fequation})  and
obtain the approximate equation
\begin{eqnarray}\label{Faequation}
&&\partial_{\zeta} F_{(a)}(\zeta,t)+i\beta L\partial^2_{t}F_{(a)}(\zeta,t)-i\mu(t)\left(F_{(a)}(\zeta,t)+\overline{F}_{(a)}(\zeta,t)\right) \approx {\eta}_{(a)}(\zeta,t)-\nonumber \\&& 2 i \beta L \Big(\partial_{t}\theta_{0}(z,\,t)\Big) \partial_{t}F_{(a)}(\zeta,t)+  2 i \gamma \rho(t) L
\left[ 2F_{(a)}(\zeta,t) Re[\widetilde{\Phi}^{(1)}(\zeta,t)]+\overline{F}_{(a)}(\zeta,t)\widetilde{\Phi}^{(1)}(\zeta,t) \right]- \nonumber \\&& i \beta L  F_{(a)}(\zeta,t) \left[i \Big(\partial^2_{t}\theta_{0}(z,\,t)\Big)- \Big(\partial_{t}\theta_{0}(z,\,t)\Big)^2  \right] \,,
\end{eqnarray}
where $\zeta=z/L$, $\theta_{0}(z,\,t)=\phi_{0}(t)+\mu(t) \zeta$ and we have used that for a slowly varying function $s(t)$ with the bandwidth $W \ll W_a$ one has at least with accuracy ${\cal O}(\tau_a/\widetilde{\delta t})$:
\begin{eqnarray}\label{mainapproximation}
s_{(a)}(t) \approx s(t), \quad \int^{t+\tau_a}_{t-\tau_a}\frac{dt'}{2\tau_a} F(z,t') s(t') \approx s(t) F_{(a)}(z,t), \quad \int^{t+\tau_a}_{t-\tau_a}\frac{dt'}{2\tau_a}\Big[ \partial^2_{t'} F(z,t')\Big] \approx  \partial^2_{t} F_{(a)}(z,t).
\end{eqnarray}
It is worth noting that from the practical and numerical points of view the averaging (frequency filtering) of the detector can affect on transmitted information (distort the output signal) much greater than both noise effects and effects of small dispersion. Therefore to guarantee the validity of our calculation method we should keep strict watch for hierarchy of our approximations when  linearizing, see Eq.~(\ref{Fequation}), and when obtaining Eq.~(\ref{Faequation}) from (\ref{taurelation}) and (\ref{mainapproximation}). It means that for appropriateness of the derivation of Eq.~(\ref{Faequation}) we require, e.g., for the first approximation in Eq.~(\ref{mainapproximation}), that $\max|(s_{(a)}(t) - s(t))/(s_{(a)}(t) + s(t))| \ll \sqrt{1/\mathrm{SNR}}, \widetilde{\beta} $ and so on.

For the averaged and multiplied by L noise ${\eta}_{(a)}(z,t)=L\int^{t+\tau_a}_{t-\tau_a}\frac{dt'}{2\tau_a} \widetilde{\eta}(z,t')$ we can obtain from Eq.~ (\ref{noisecorrelatort}) the following statistics:
\begin{eqnarray}\label{etaastatistics1}
&& \langle \eta_{(a)} (\zeta,t)\rangle_{\eta}=0, \qquad \langle \eta_{(a)} (\zeta,t)\bar{\eta}_{(a)}(\zeta^\prime,t^\prime)\rangle_{\eta} =  Q L
\delta(\zeta-\zeta^\prime) {K}_{a}(t,t^\prime)\,,
\end{eqnarray}
where our averaging procedure (\ref{OSaveraging}) results in the following form of the function ${K}_{a}(t,t^\prime)$:
\begin{eqnarray}\label{etaastatistics2}
&& {K}_{a}(t,t^\prime)=K(t-t')=\frac{1}{2\tau_a} \Big(1-\frac{|t-t'|}{2\tau_a}\Big)\theta\left(|t-t'|\leq 2 \tau_a\right),
\end{eqnarray}
and for the frequency domain we have $\langle \eta_{(a)}
(\zeta,\omega)\bar{\eta}_{(a)}(\zeta^\prime,\omega^\prime)\rangle_{\eta}
=  Q L \delta(\zeta-\zeta^\prime)\hat{K}_{a}(\omega,\omega^\prime)$,
where  for our  procedure (\ref{OSaveraging})
$\hat{K}_{a}(\omega,\omega^\prime)=2 \pi
\delta(\omega-\omega^\prime) K_{\omega}$,
$K_{\omega}=\mathrm{sinc}^2(\omega \tau_a)$. Note that for the limit
$\tau_a \rightarrow 0$ (from relation (\ref{taurelation}) one can
see that this limit is acceptable for dispersionless case) $K(t-t')
\rightarrow \delta(t-t')$, and $K_{\omega} \rightarrow 1$.

%==================
\subsection{Solution of the averaged equation (\ref{Faequation})}
Note that the time scale of the solution $F_{(a)}(\zeta,t)$ and 
the first term (${\eta}_{(a)}(\zeta,t)$) in the r.h.s. of equation
(\ref{Faequation}) is $\tau_a$. The second term in the r.h.s. of
(\ref{Faequation}) is of order of $\widetilde{\beta}_{a} \ll 1$
(since the time scale of $\theta_{0}(z,\,t)$ is $\widetilde{\delta
t}$). The other terms in the r.h.s. of  (\ref{Faequation}) are of
order of $\widetilde{\beta} \ll \widetilde{\beta}_{a}$ (since there
are no derivatives of $F_{(a)}(\zeta,t)$ in these terms). This means
that we can solve equation (\ref{Faequation}) perturbatively in
small parameter $\widetilde{\beta}_{a} \ll 1$. To find the solution
we use the Laplace transformation over $\zeta=z/L$ and  the Fourier
transformation over ``fast'' time, i.e., all slowly varying
functions $\mu(t)$ and $\phi_0(t)$ are treated as constants
(``freezing'' of the coefficients), but then their time derivatives
emerge as corrections in  $\widetilde{\beta}_{a}$:
\begin{eqnarray}
F_{(a)}(P,\omega;\{\mu(t),\phi_0(t)\})=\int^{\infty}_{0}{d\zeta}e^{-P
\zeta} \int dt' e^{i\omega t'}
F_{(a)}(\zeta,t';\{\mu(t),\phi_0(t)\}).
\end{eqnarray}
Here we explicitly emphasize $\mu(t)$ and $\phi_0(t)$--dependance of
our solution. We will omit this dependance in what follows. To find
$F_{(a)}(P,\omega)$ and $\overline{F}_{(a)}(P,-\omega)$ we have to solve the
following algebraic system resulting from Eq.~(\ref{Faequation}):
\begin{eqnarray}\label{FaequationPomega}
 P F_{(a)}(P,\omega) -  i \beta L \omega^2 F_{(a)}(P,\omega) - i \mu(t) [F_{(a)}(P,\omega)+\overline{F}_{(a)}(P,-\omega)]&=&R[P,\omega], \nonumber \\
P \overline{F}_{(a)}(P,-\omega) + \beta L \omega^2
\overline{F}_{(a)}(P,-\omega) + i \mu(t)
[F_{(a)}(P,\omega)+\overline{F}_{(a)}(P,-\omega)] &=&
\overline{R}(P,-\omega),
\end{eqnarray}
where in the leading (zero) order in $\widetilde{\beta}_{a}$ we have from the r.h.s. of Eq.~(\ref{Faequation})
\begin{eqnarray}\label{R0}
R^{(0)}[P,\omega]=\eta_{(a)} (P,\omega), \qquad
\overline{R}^{(0)}[P,-\omega]=\overline{\eta}_{(a)} (P,-\omega).
\end{eqnarray}
And we easily find the leading order contribution
$F^{(0)}_{(a)}(P,\omega;\{\mu, \phi_0\})$ from
(\ref{FaequationPomega}). The solution
$F^{(0)}_{(a)}(P,\omega;\{\mu, \phi_0\})$ is the even function of
$\omega$ and it depends on $\mu(t)$ only.

In the next-to-leading order in $\widetilde{\beta}_{a}$ (i.e., order $\widetilde{\beta}^{1}_{a}$) we have the
contributions from the second term in r.h.s. of
Eq.~(\ref{Faequation}) and the contributions from the second time
derivatives of the leading order, see the second term in the r.h.s. of
the following equality
\begin{eqnarray}
\!\!\!\frac{d^2}{dt^2} F_{(a)}(\zeta,t;\{s(t)\})=
\frac{\partial^2}{\partial t^2}F_{(a)}(\zeta,t;\{s\})+2 \dot{s}
\frac{\partial^2}{\partial s
\partial
t}F_{(a)}(\zeta,t;\{s\})+\Big(\ddot{s}\frac{\partial}{\partial s}+
\dot{s}^2 \frac{\partial^2}{\partial
s^2}\Big)F_{(a)}(\zeta,t;\{s\}).
\end{eqnarray}
Thus we have
\begin{eqnarray}\label{R1}
&&R^{(1)}[P,\omega]=-2 \beta L \omega \Big[\dot{\mu}
\frac{\partial}{\partial \mu}+i \left(\dot{\phi_0} -
\dot{\mu}\frac{\partial}{\partial
P}\right)\Big]F^{(0)}_{(a)}(P,\omega;\{\mu, \phi_0\}), \nonumber
\\&& \overline{R}^{(1)}[P,-\omega]=2 \beta L \omega \Big[\dot{\mu}
\frac{\partial}{\partial \mu}-i \left(\dot{\phi_0} - \dot{\mu}
\frac{\partial}{\partial
P}\right)\Big]\overline{F}^{(0)}_{(a)}(P,-\omega;\{\mu, \phi_0\}).
\end{eqnarray}
Note that solution $F^{(1)}_{(a)}(P,\omega;\{\mu, \phi_0\})$ of Eq.
(\ref{FaequationPomega}) with r.h.s. (\ref{R1}) is the odd function
of $\omega$.

In the next-to-next-to-leading order in $\widetilde{\beta}_{a}$ (i.e., in the order
$\widetilde{\beta}^2_{a} \sim \widetilde{\beta}$) we have two different sources for these corrections: the
second order corrections in parameter $\widetilde{\beta}_{a}$  and
the first order corrections in parameter $\widetilde{\beta}=\beta L
/\widetilde{\delta t}^2 $. However, we can separate these types
of corrections in the following way:
$R^{(2)}[P,\omega]=R^{(2.1)}[P,\omega]+R^{(2.2)}[P,\omega]$.  The
r.h.s. resulting in corrections of the first order in
$\widetilde{\beta}$ reads
\begin{eqnarray}\label{R21}
&&R^{(2.1)}[P,\omega]=-i \beta L\left[ i
\Big(\ddot{\phi_{0}}-\ddot{\mu}\frac{\partial}{\partial
P}\Big)-\Big(\dot{\phi_{0}}-\dot{\mu}\frac{\partial}{\partial
P}\Big)^2 + \ddot{\mu} \frac{\partial}{\partial \mu}+ \dot{\mu}^2
\frac{\partial^2}{\partial \mu^2}\right]
F^{(0)}_{(a)}(P,\omega;\{\mu, \phi_0\}) + \nonumber \\&& 2 i
\frac{\mu}{\rho}\left\{2 \mathrm{Re} \widetilde{\Phi}^{(1)}\Big(-L
\frac{\partial}{\partial P},t\Big)F^{(0)}_{(a)}(P,\omega;\{\mu,
\phi_0\})+ \widetilde{\Phi}^{(1)}\Big(-L \frac{\partial}{\partial
P},t\Big)\overline{F}^{(0)}_{(a)}(P,-\omega;\{\mu, \phi_0\})
\right\}.
\end{eqnarray}
The Laplace transformation replaces $z$ by the operator
$-L\frac{\partial}{\partial P}$. It is obvious that
$R^{(2.1)}[P,\omega]$ is proportional to $\beta L \ddot{s} \sim
\widetilde{\beta}$ or $\beta L \dot{s}^2 \sim \widetilde{\beta}$,
where $s$ is slowly varying function ($\mu(t)$, $\phi_0 (t)$). The
r.h.s. of the system (\ref{FaequationPomega}) resulting in
corrections of the second order in $\widetilde{\beta}_{a}$ has the
same structure as (\ref{R1}): $R^{(2.2)}[P,\omega]=-2 \beta L \omega
\Big[\sum_{s=\{\mu,\phi_0,\dot{\mu},\dot{\phi_0}\}}\dot{s}
\frac{\partial}{\partial s}+i \left(\dot{\phi_0} -
\dot{\mu}\frac{\partial}{\partial
P}\right)\Big]F^{(1)}_{(a)}(P,\omega;\{\mu, \phi_0\})$. Fortunately, these corrections ($\propto \widetilde{\beta}^2_{a}$) are not essential for the further calculations.

Now we present the result of our calculation in the first order in
the noise $\eta_{(a)}$ and in the leading and next-to-leading order
in $\widetilde{\beta}$ (i.e., we take into account $\widetilde{\beta}^{0}_a$, $\widetilde{\beta}_a$, and $\widetilde{\beta}_a^2$ contributions):
\begin{eqnarray}\label{Fazt}
\!\!F_{(a)}(\zeta,t)=\int \frac{d\omega}{2\pi}e^{-i \omega t}\!\!\!\int^{\zeta}_{0}\!\!d\xi\Big[\eta_{(a)} (\xi,\omega) M_{\eta}(\zeta,\xi,\omega;\mu,\phi_0)+\bar{\eta}_{(a)}(\xi,-\omega) M_{\bar{\eta}}(\zeta,\xi,\omega;\mu,\phi_0)\Big],
\end{eqnarray}
\begin{eqnarray}\label{barFazt}
\!\!\overline{F}_{(a)}(\zeta,t)=\int \frac{d\omega}{2\pi}e^{-i \omega t}\!\!\!\int^{\zeta}_{0}\!\!d\xi\Big[\eta_{(a)} (\xi,\omega) \overline{M}_{\bar{\eta}}(\zeta,\xi,-\omega;\mu,\phi_0)+\bar{\eta}_{(a)}(\xi,-\omega) \overline{M}_{{\eta}}(\zeta,\xi,-\omega;\mu,\phi_0)\Big].
\end{eqnarray}
In the further consideration we need only the the expansion of $M_{\eta}$ and $M_{\bar{\eta}}$
in $\beta L \omega^2$. The reason is the subsequent projection of these functions $M_{\eta}$ and $M_{\bar{\eta}}$ on the basis (\ref{Xtmodel}) with the frequency support $W$: it reduces all powers of $\beta L \omega^2$ to powers of $\beta L W^2 \propto \widetilde{\beta}$, see the detailed explanation in Sec. \ref{SecPCC}.

In the leading order in $\widetilde{\beta}_{a}$ we have the following result presented as the expansion in $\beta L \omega^2$:
\begin{eqnarray}\label{MLOsmallbeta1}
M^{(0)}_{\eta}(\zeta,\xi,\omega;\mu(t),\phi_0(t))\approx1+ i \mu(t)(\zeta-\xi)+\beta L \omega^2\Big(i(\zeta-\xi)-\mu(t)(\zeta-\xi)^2-\frac{i \mu(t)^2}{3}(\zeta-\xi)^3\Big),
\end{eqnarray}
\begin{eqnarray}\label{MLOsmallbeta2}
M^{(0)}_{\bar{\eta}}(\zeta,\xi,\omega;\mu(t),\phi_0(t))\approx i  \mu(t)(\zeta-\xi) -\beta L \omega^2 \frac{i \mu(t)^2}{3}(\zeta-\xi)^3.
\end{eqnarray}
%where $\nu(\omega,\mu)=\sqrt{\beta L \omega^2(2 \mu+\beta L \omega^2)}$. For the brevity sake we will omit the arguments of $\nu(\omega,\mu)$, $\mu(t)$, and $\phi_{0}(t)$.

The first correction in $\widetilde{\beta}_{a}$ contains the terms proportional to $\beta L \omega \dot{\mu}$ and $\beta L \omega \dot{\phi_0}$:
\begin{eqnarray}\label{MNLOsmallbeta1}
M^{(1)}_{\eta}(\zeta,\xi,\omega;\mu(t),\phi_0(t))\approx \frac{\beta L \omega}{3}(\zeta -\xi )\Big\{6 \dot{\phi_{0}}\Big[ (\zeta -\xi )\mu-i \Big]+\dot{\mu}\Big[(\zeta -\xi )(5\zeta +\xi)\mu-6 i \zeta\Big] \Big\},
\end{eqnarray}
\begin{eqnarray}\label{MNLOsmallbeta2}
M^{(1)}_{\bar{\eta}}(\zeta,\xi,\omega;\mu(t),\phi_0(t))\approx\frac{\beta L \omega}{3}(\zeta -\xi )^2 \Big\{ 6 \dot{\phi_{0}} \mu + \dot{\mu}\Big[ (5\zeta +\xi)\mu-3 i\Big]\Big\}.
\end{eqnarray}

Finally, we found the first corrections in parameter $\widetilde{\beta}$, i.e., $M^{(2.1)}_{\eta}(\zeta,\xi,\omega;\mu(t),\phi_0(t))$ and $M^{(2.1)}_{\bar{\eta}}(\zeta,\xi,\omega;\mu(t),\phi_0(t))$. Here we present the expansion of this result in parameter $\beta L \omega^2$:
\begin{eqnarray}\label{MNNLOsmallbeta1}
&& M^{(2.1)}_{\eta}(\zeta,\xi,\omega;\mu(t),\phi_0(t))\approx -\beta L (\zeta -\xi )\Big\{ \mu \frac{\dot{\rho}^2}{3\rho^2}\Big( 10 \mu ^2 \left(\zeta ^3-\xi ^3\right)-3 i \mu  \left(7 \zeta ^2+4 \zeta  \xi +5 \xi
   ^2\right)- \nonumber \\&& 6 \zeta\Big)+  \frac{2 \mu (\zeta +\xi)}{\rho}\dot{\rho} \dot{\phi_0} \Big(2 \mu  (\zeta -\xi )-3 i\Big)+  \dot{\phi_0}^2 \Big( \mu  (\zeta -\xi )-i \Big)+\ddot{\phi_0} \Big(\mu ^2 \left(\zeta ^2-\xi ^2\right)-\nonumber \\&&i \mu  (3 \zeta +\xi )- 1  \Big) +
   \frac{2\mu \ddot{\rho}}{3\rho}\Big(\mu ^2 \left(\zeta ^3-\xi ^3\right)-i \mu  \left(4 \zeta ^2+\zeta  \xi +\xi
   ^2\right)-3 \zeta\Big) \Big\}.
\end{eqnarray}
\begin{eqnarray}\label{MNNLOsmallbeta2}
&& M^{(2.1)}_{\bar{\eta}}(\zeta,\xi,\omega;\mu(t),\phi_0(t))\approx -\beta L (\zeta -\xi )\Big\{ \mu \frac{\dot{\rho}^2}{3\rho^2}\Big( 10 \mu ^2 \left(\zeta ^3+\xi ^3\right)-i \mu  \left(11 \zeta ^2+2 \zeta  \xi +5 \xi
   ^2\right)- \nonumber \\&& 3 \zeta +3 \xi \Big) + \frac{2 \mu}{\rho}\dot{\rho} \dot{\phi_0}  \Big(2 \mu  \left(\zeta ^2+\xi ^2\right)-i (\zeta +\xi ) \Big) +  \dot{\phi_0}^2 \mu (\zeta +\xi)+ \ddot{\phi_0} \mu\Big(\mu  \left(\zeta ^2+\xi^2 \right)-2 i \zeta\Big)+\nonumber \\&& \frac{2\mu \ddot{\rho}}{3\rho}\Big(\mu ^2 \left(\zeta ^3+\xi ^3\right)-3 i \zeta ^2 \mu -3 \zeta\Big) \Big\}.
\end{eqnarray}

Note that the function $F_{(a)}(\zeta,t)$ presented in the unexpanded form allows one to calculate the averaged  output signal, i.e., $\Psi_{(a)}(L,t)=\Phi(L,t)+F_{(a)}(\zeta=1,t) e^{i \phi_{0}(t)+i \mu(t)} $,  and then to find all correlators of the output signal.

To summarize, we have performed the following procedures:
\begin{eqnarray}
\underset{bandwidth:\,  W=2\pi/\widetilde{\delta t}}{X(t)}
\xrightarrow[Eq.~(\ref{startingCannelEqt})]{{NLSE\,with\,noise}} \psi(L,t) \underset{Detector}{\underbrace{\xrightarrow[Eq.~(\ref{psi})]{e^{-i\theta_0}(\psi-\Phi)}
\underset{W'\gg
W}{F(\zeta=1,t)}\xrightarrow[Eq.\,(\ref{OSaveraging})]{{Averaging\,
with\, \tau_a}}}}   \underset{
W_a=2\pi/\tau_a}{F_{(a)}(\zeta=1,t)} .\nonumber
\end{eqnarray}
Let us emphasize that the particular averaging procedure
(\ref{OSaveraging}) is not crucial for our consideration: we require
only the inequality (\ref{taurelation}) and the important
approximations (\ref{mainapproximation}) to hold true.

%===================================
\section{Input signal recovery: backward NLSE propagation}
\label{SecInpusignal}

The function $F_{(a)}(\zeta,t)$ describes the noise impact in our channel when we measure the output signal filtered with the bandwidth $W_a=2\pi/\tau_a$.
Now we describe the procedure of the input signal recovery from the output signal. This procedure is not unique and depends on the output signal filter and the input signal model, see Sec. \ref{SecInput signal}.

On the base of the measured (with filtering process described above) output signal $\Psi_{(a)}(L,t)=\Phi(L,t)+F_{(a)}(\zeta=1,t) e^{i \phi_{0}(t)+i \mu(t)} $ we can restore the input signal $X(t)$ by the backward NLSE evolution. We denote this analytically restored signal as $\widetilde{X}(t)=\hat{L}^{-1}[\Psi_{(a)}(L,t)]$, where $\hat{L}^{-1}$ is the inverted  evolution operator of the NLSE (\ref{startingCannelEqt}) with zero noise (backward NLSE propagation). In the case of small $\beta$ we know the explicit form of this operator in the perturbation theory. We present the restored signal in the form
$\widetilde{X}(t)=X(t)+\delta \widetilde{X}(t)$, and for $\beta \neq 0$  one has in the leading ($\widetilde{\beta}^0$) and in next-to-leading ($\widetilde{\beta}^1$) orders:
\begin{eqnarray}\label{deltaX}
&& \delta \widetilde{X}(t)\approx  e^{i \phi_0(t)}\Big[(1-i\mu(t))
F_{(a)}-i \mu(t)\overline{F}_{(a)}\Big]+\nonumber \\&& \beta L e^{i
\phi_0(t)}\Big\{{F}_{(a)} a_{1}(t)+\overline{F}_{(a)}
a_{2}(t)+{F}'_{(a)} a_{3}(t)+\overline{F}'_{(a)}
a_{4}(t)+{F}''_{(a)} a_{5}(t)+\overline{F}''_{(a)} a_{6}(t)\Big\},
\end{eqnarray}
where coefficients $a_k(t)$ have the form:
\begin{eqnarray}\label{a1a6}
&&a_{1}(t)=-\ddot{\phi_{0}}-i \dot{\phi_{0}}^2-\mu  \left(\frac{4 i \dot{\rho} \dot{\phi_{0}}}{\rho }+i \ddot{\phi_{0}}+\dot{\phi_{0}}^2\right)-\frac{\mu ^2 }{3 \rho^2}\left(2 i \rho  \ddot{\rho}+12 \rho  \dot{\rho} \dot{\phi_{0}}+11 i \dot{\rho}^2+3 \rho^2
   \ddot{\phi_{0}}\right)-\nonumber \\&&\frac{2 \mu^3 }{3 \rho^2}\left(\rho \ddot{\rho}+5 \dot{\rho}^2\right), \qquad
a_{2}(t)=\mu  \left(\frac{\dot{\rho}^2}{\rho ^2}+\dot{\phi_{0}}^2\right)+\mu^2
   \left(\frac{4 \dot{\rho} \dot{\phi_{0}}}{\rho }-\frac{i \dot{\rho}^2}{3 \rho
   ^2}+\ddot{\phi_{0}}\right)+\frac{2 \mu ^3 }{3 \rho^2}\left(\rho  \ddot{\rho}+5 \dot{\rho}^2\right), \nonumber \\&&   a_{3}(t)=-2 \dot{\phi_{0}}+2 i \mu  \dot{\phi_{0}}+\frac{2 i \mu ^2 \dot{\rho}}{3 \rho },
a_{4}(t)=\mu  \left(\frac{2 \dot{\rho}}{\rho }+2 i \dot{\phi_{0}}\right)+\frac{2 i \mu
   ^2 \dot{\rho}}{3 \rho }, \nonumber \\&&  a_{5}(t)=i+\mu-\frac{1}{3} i \mu^2 , \qquad a_{6}(t)=-\frac{1}{3} i \mu^2.
\end{eqnarray}
Note that in Eq.~(\ref{deltaX}) we have retained only linear in $F_{(a)}$ terms: it is correct if $\mathrm{SNR}^{-1} \ll \widetilde{\beta} \ll 1$.
In the first line of the Eq.~(\ref{deltaX}) we should take $F_{(a)}$ and $\overline{F}_{(a)}$ in the leading, i.e., $\widetilde{\beta}^{\,0}$, see Eqs.~(\ref{MLOsmallbeta1}), (\ref{MLOsmallbeta2})  and in the next-to-leading ($\widetilde{\beta}^{\,1}$) orders: see Eqs. (\ref{MNLOsmallbeta1}), (\ref{MNLOsmallbeta2}). In the second line $F_{(a)}$, $\overline{F}_{(a)}$  and their derivatives are taken in the leading order only, see Eqs.~(\ref{MLOsmallbeta1}), (\ref{MLOsmallbeta2}).
Let us present now the function $\delta \widetilde{X}(t)$ in the explicit form:
\begin{eqnarray}\label{deltaXexplicit1}
\delta \widetilde{X}(t)=e^{i \phi_0(t)} \int \frac{d \omega}{2 \pi} e^{-i \omega t} \int^{1}_{0} d\xi\Big[\eta_{(a)} (\xi,\omega) M_{(x)\eta}(\xi,\omega;\mu,\phi_0)+\bar{\eta}_{(a)}(\xi,-\omega) M_{(x)\bar{\eta}}(\xi,\omega;\mu,\phi_0)\Big],
\end{eqnarray}
where in our consideration we need only the expansion of the coefficients  $M^{(x)}_{\eta}$ and $M^{(x)}_{\bar{\eta}}$ in $\beta$ up to the first order corrections:
\begin{eqnarray}\label{Mx}
&& M_{(x)\eta}(\xi,\omega;\mu,\phi_0)=1-i \mu  \xi +\beta  L \omega  \Big\{\frac{2 \mu ^2 \xi ^3 \dot{\rho}}{3 \rho }+\left(2 \mu  \xi ^2+2 i \xi \right)
   \dot{\phi_{0}}\Big\}+\beta  L \omega ^2 \Big\{\frac{1}{3} i \mu ^2 \xi ^3-\mu  \xi ^2-i \xi \Big\}+\nonumber \\&&\beta
    L \Big\{ \frac{\dot{\rho} \dot{\phi_{0}}}{\rho} \left(-{4 \mu ^2 \xi ^3}-{2 i \mu  \left(3 \xi ^2-2 \xi
   +1\right)}\right)+\left(-\mu ^2 \xi ^3-i \mu  \xi ^2-\xi \right) \ddot{\phi_{0}}+\frac{\ddot{\rho}}{3 \rho}
   \left(-{2 \mu ^3 \xi ^4}-{2 i \mu ^2 \xi ^3}\right)+\nonumber \\&& \frac{\dot{\rho}^2}{3 \rho ^2}
   \left(-{10 \mu ^3 \xi ^4}-{i \mu ^2 \left(15 \xi ^3-12 \xi +8\right)}\right)+\left(-\mu  \xi ^2-i \xi \right) \dot{\phi_{0}}^2\Big\}+{\cal O}(\beta^2),
\end{eqnarray}
\begin{eqnarray}\label{Mxbar}
&&M_{(x)\bar{\eta}}(\xi,\omega;\mu,\phi_0)=-i \mu  \xi + \beta L\omega ^2  \frac{i \mu ^2  \xi ^3}{3} +\beta  L \omega  \Big\{2 \mu  \xi ^2 \dot{\phi_{0}}+\frac{2 \dot{\rho}}{3 \rho} \left({ \mu ^2 \xi ^3}-{ 3 i \mu  \xi ^2}\right)\Big\}+ \nonumber \\&& \beta  L
   \Big\{\frac{2 \mu ^3 \xi ^4 \ddot{\rho}}{3 \rho }+\frac{\dot{\rho}\dot{\phi_{0}} }{\rho}\left({4 \mu ^2 \xi ^3}-{2 i \mu  (\xi -1)^2}\right)+\mu ^2 \xi ^3 \ddot{\phi_{0}}+\frac{\dot{\rho}^2}{3 \rho ^2}
   \Big({10 \mu ^3 \xi ^4}- {i \mu ^2 \left(5 \xi ^3-12 \xi +8\right)}+\nonumber \\&& 3{\mu  \xi ^2}\Big)+ \mu  \xi ^2 \dot{\phi_{0}}^2\Big\}+{\cal O}(\beta^2).
\end{eqnarray}

%===================================
\section{Correlators and conditional probability density function $P[\{\widetilde{C}\}|\{C\}]$ }
\label{SecPCC}

Now we can calculate the  projection of the recovered function
$\widetilde{X}(t)=X(t)+\delta \widetilde{X}(t)$ on the basis
$g_{k}(t)$:
\begin{eqnarray} \label{projectiontilde}
\widetilde{C}_k=\int_T \frac{dt}{T_0} g_k(t) \widetilde{X}(t)=C_k + \delta\widetilde{C}_k, \qquad \delta\widetilde{C}_k=\int_T \frac{dt}{T_0} g_k(t) \delta \widetilde{X}(t),
\end{eqnarray}
where $C_k$ are known coefficients: they are determined by the equation (\ref{Ckrecover}) on the base of the input signal $X(t)$, and corrections $\delta\widetilde{C}_k$ are responsible for the noise impact in the channel, see Eq.~(\ref{deltaX}). We can calculate $\delta\widetilde{C}_k$ using Eqs. (\ref{Mx}), (\ref{Mxbar}). Let us note that in the representation (\ref{deltaXexplicit1}) there is the  integral over frequencies with the maximal frequency of order of $W_a=2\pi/\tau_a$. Indeed,  this maximal frequency is determined by the function $K_{\omega}$ in the statistics  of the averaged noise $\eta_{(a)} (\xi,\omega)$. However, when we calculate the projections (\ref{projectiontilde}) we cut off all frequencies above $W=2\pi/{\widetilde{\delta t}}$. That is to say, these projections (\ref{projectiontilde}) play the frequency filtering role, and  under the subsequent frequency integration all parameters $\beta L \dot{s}^2$,  $\beta L \ddot{s}$, $\beta L \omega \dot{s}$, and $\beta L \omega^2$ become of the same order $\widetilde {\beta}=\beta L/\widetilde{\delta t}^2$.

With our accuracy, i.e.,  neglecting the overlapping of the base functions $g_{k}(t)$, it is easy to obtain the following correlators in the leading and next-to leading order in  $\widetilde{\beta}=\beta L /\widetilde{\delta t}^2$:
%\begin{eqnarray}\label{ak1ak2andak1barak2-1}
%\langle \delta \widetilde{C}_{k_1} \delta \widetilde{C}_{k_2} \rangle_{\eta}\approx  {{
%\delta_{k_1,\,k_2} e^{2 i \phi_{k_1}} \frac{ Q L}{\widetilde{\delta t}}  \Big( -\frac{i \widetilde{\gamma} r^2_{k_1} }{\sqrt{2 \pi }}-\frac{2 \widetilde{\gamma}^2 r^4_{k_1} N_t}{3 \sqrt{3}
%   \pi }}}  {{{+  \frac{4 \widetilde{\gamma}^3 r^6_{k_1} N_t^2}{15 \pi ^{3/2}}\,\widetilde{\beta }}\Big)}},
%\end{eqnarray}
%\begin{eqnarray}
%\label{ak1ak2andak1barak2-2}
%\langle \delta \widetilde{C}_{k_1} \overline{\delta \widetilde{C}}_{k_2} \rangle_{\eta}\approx
%\delta_{k_1,\,k_2}   \frac{ Q L}{\widetilde{\delta t}}
%\Big(\frac{1}{N_t} +\frac{2 \widetilde{\gamma}^2 r^4_{k_1}  N_t}{3
%\sqrt{3} \pi } -\frac{4
%\widetilde{\gamma}^3 r^6_{k_1}   N_t^2}{15 \pi
%^{3/2}}\widetilde{\beta }\Big),
%\end{eqnarray}
%\begin{eqnarray}\label{ak1ak2andak1barak2-3}
%\langle \delta \widetilde{C}_{k} \rangle_{\eta} = {\cal O}\left(Q L W'\right),
%\end{eqnarray}
\begin{eqnarray}\label{ak1ak2andak1barak2-1}
\langle \delta \widetilde{C}_{k_1} \delta \widetilde{C}_{k_2}
\rangle_{\eta}\approx \delta_{k_1,\,k_2} \frac{ Q L}{T_0}  \gamma L
{C}^2_{k_1}  \Big(-i n_4 -\frac{2}{3}\gamma L |C_{k_1}|^2 n_6
+\widetilde{\beta } \frac{8}{15} \gamma^2 L^2 |C_{k_1}|^4 n_8
\Big),
\end{eqnarray}
\begin{eqnarray}
\label{ak1ak2andak1barak2-2} \langle \delta \widetilde{C}_{k_1}
\overline{\delta \widetilde{C}}_{k_2} \rangle_{\eta}\approx
\delta_{k_1,\,k_2}   \frac{ Q L}{T_0} \Big(1+\frac{2}{3}\gamma^2 L^2
|C_{k_1}|^4 n_6 -\widetilde{\beta }\frac{8}{15} \gamma^3 L^3
|C_{k_1}|^6 n_8
 \Big),
\end{eqnarray}
\begin{eqnarray}\label{ak1ak2andak1barak2-3}
\langle \delta \widetilde{C}_{k} \rangle_{\eta} =-i {C}_{k} \frac{Q
L}{\Delta}\gamma L \left(1-i \frac{\gamma L |C_{k}|^2 n_4}{3}
\right)  +{\cal O}\left(\frac{Q L}{\Delta}  \widetilde{\beta
}\right),
\end{eqnarray}
here ${ Q L}/{\Delta}=Q L W'/(2\pi)$ is the  noise power in the
whole noise bandwidth $W'$; and we have introduced the integrals $n_s=\int_T\dfrac{dt}{T_0} g^s_0(t)$: $n_2=1$,
$n_4 = \frac{N_t}{\sqrt{2\pi}}$, $n_6 = \frac{N_t^2}{\pi\sqrt{3}}$,
$n_8 = \frac{N_t^3}{2\pi\sqrt{\pi}}$, where the sparsity
$N_{t}=T_0/\widetilde{\delta t} \gg 1$. These simple forms of
$\widetilde{\beta }$-corrections in Eqs.~
(\ref{ak1ak2andak1barak2-1}), (\ref{ak1ak2andak1barak2-2}) is the
particular feature of the Gaussian pulses (\ref{Xtmodel}). Terms devoid of dispersion $\widetilde{\beta }$ in
Eqs.~(\ref{ak1ak2andak1barak2-1})--(\ref{ak1ak2andak1barak2-3}) are
in agreement with the results of our paper \cite{Terekhov:2018}, see
Eqs. (19)--(21) therein. Note that we have not calculated the
corrections in $\widetilde{\beta }$ for the correlator
(\ref{ak1ak2andak1barak2-3}) since when solving
Eq.~(\ref{Faequation}) we have taken into account only linear in
noise $\eta(z,t)$ terms. However these corrections in the correlator
(\ref{ak1ak2andak1barak2-3}) are not essential for the informational
channel characteristics evaluated in the leading order in
$1/\mathrm{SNR}$, see the discussion in \cite{Terekhov:2018}. When obtaining (\ref{ak1ak2andak1barak2-1}) and
(\ref{ak1ak2andak1barak2-2}) we have used that
$K_{\omega=W}=\mathrm{sinc}^2(W \tau_a) \approx 1$ and the details
of the output signal filtering become irrelevant if ${\tau^2_a}/{\widetilde{\delta t}}^2 \ll \widetilde{\beta }$ (i.e., the last inequality in Eq.~(\ref{taurelation}) is strengthened as $\beta L/\tau^2_a  \gg 1$). Also we have
neglected the overlapping effects (of order of $\exp[- c \,N^2_t]$)
and derivatives of phases.

Two correlators (\ref{ak1ak2andak1barak2-1}),
(\ref{ak1ak2andak1barak2-2})  are sufficient for calculation of the
conditional probability density function in the leading order in
parameter $1/\mathrm{SNR}$:
\begin{eqnarray}\label{PCC}
P[\{\widetilde{C}\}|\{C\}]=\prod^{N}_{m=-N}P_{m}[\widetilde{C}_m|C_m],
\end{eqnarray}
\begin{eqnarray}\label{PmCC}
P_{m}[\widetilde{C}_m|C_m]=P^{(0)}_{m}[\widetilde{C}_m|C_m]\left(1+\widetilde{\beta
}\frac{8 \mu^3_{m} n_8 \, }{15 (1+\xi^2 \mu^2_{m}/3)}\Big[1-2 T_0
\frac{(y_m+ n_4 \mu_m x_m )^2}{Q L(1+\xi^2 \mu^2_{m}/3)}\Big]+{\cal
O}(\widetilde{\beta }^2)\right),
\end{eqnarray}
here $P^{(0)}_{m}[\widetilde{C}_m|C_m]$ was found in
Ref.~\cite{Terekhov:2018} in the nondispersive channel model:
\begin{eqnarray}\label{P0mCC}
P^{(0)}_{m}[\widetilde{C}_m|C_m] \approx  \frac{T_0}{\pi Q L
\sqrt{1+\xi^2 \mu^2_{m}/3}} \exp\left[- T_0  \frac{\left(1+4n_6
\mu^2_m /(3)\right)x^2_m+2 x_m y_m \mu_m n_4  +y^2_m}{Q L
\left(1+\xi^2 \mu^2_{m}/3\right)}\right],
\end{eqnarray}
where we have introduced the following  notations inherent for the
nondispersive model: $x_m= \mathrm{Re} \left[e^{-i \phi_m} \Big\{\delta
\widetilde{C}_m-\langle \delta \widetilde{C}_{m} \rangle_{\eta}
\Big)\Big\}\right]$, $y_m= \mathrm{Im} \left[e^{-i \phi_m}
\Big\{\delta \widetilde{C}_m-\langle \delta \widetilde{C}_{m}
\rangle_{\eta} \Big)\Big\}\right]$, $\phi_m=\arg {C}_m$, $\mu_m=\gamma L |C_m|^2$, $\xi^2=(4n_6 - 3 n^2_4)$.
The parameter $\xi^2$ obeys the inequality  $\xi^2>n_6 >0$ due to
Cauchy-Schwarz-Buniakowski inequality. Following the technique of Ref.~\cite{Terekhov:2017}  for given
$\widetilde{\beta }$-corrections in the representation of
$P[\{\widetilde{C}\}|\{C\}]$ it is straightforward to calculate the
$\widetilde{\beta }$-corrections to all informational channel
characteristics. To summarize, this technique reads:
\begin{eqnarray}
\widetilde{C}_k\xrightarrow[]{}\underset{Eqs.\,(\ref{ak1ak2andak1barak2-1}),\,(\ref{ak1ak2andak1barak2-2})}{\mathrm{Correlators}}
\xrightarrow[]{} {P[\{\widetilde{C}\}|\{C\}]}\xrightarrow[]{}\,H[Y],\,H[Y|X]
\xrightarrow[]{} I_{P[C]} \xrightarrow[max]{}{P_{opt}[C],\,\,  C=
I_{P_{opt}[C]}}. \nonumber
\end{eqnarray}

%===================================
\section{Concluding remarks}
\label{SecConclusion}

On the basis of the stochastic NLSE with the white Gaussian noise we
consider the realistic channel model with the input signal $X(t)$ and
detector models included. For the case of large $\mathrm{SNR}$ and
small dispersion  $\widetilde{\beta }$ we present the approximate
method based on NLSE averaging with ``freezing'' of slowly varying
functions. We find the simplest two correlators of the recovered
coefficients $\widetilde{C}_k$, see Eqs. (\ref{ak1ak2andak1barak2-1}), (\ref{ak1ak2andak1barak2-2}). These correlators are base
for calculation of informational channel characteristics in the
leading order in parameter $1/\mathrm{SNR}$ and in the first order in the dispersion parameter $\widetilde{\beta }$.

\ack
%===================================
\emph{\it Acknowledgments}
All authors would like to thank the Russian Science Foundation (RSF), grant No.
16-11-10133. Also A.R. would like to thank the Russian Foundation for Basic Research (RFBR), grant No. 16-31-60031.
%Also I.S. would like to thank the Russian Science Foundation (RSF), grant No. 17-72-30006, and Ministry of Education and Science of the Russian Federation (14.Y26.31.0017).

\smallskip


\medskip

\section*{References}
\begin{thebibliography}{99}

%==================1
\bibitem{Haus:1991}
Haus H~A~ 1991  \emph{Journal of the Optical Society of America } B \textbf{8}  1122

%==================2
\bibitem{Mecozzi:1994}
Mecozzi A~ 1994, \emph{J. Lightwave Technol.} \textbf{12} 1993

%==================3
\bibitem{Iannoe:1998}
Iannoe E, Matera F, Mecozzi A and Settembre M 1998 \emph{Nonlinear  Optical Communication Networks} (New~York: John Wiley \& Sons)

%==================4
\bibitem{Turitsyn:2000}
Turitsyn S~K, Medvedev S~B, Fedoruk M~P and Turitsyna E~G 2000 \emph{Phys. Rev. } E \textbf{61} 3127

%==================5
\bibitem{MSR:1973}
Martin P C, Siggia E D and  Rose H A 1973 {\it Phys. Rev.} A {\bf 8} 423

%==================6
\bibitem{Terekhov:2014}
Terekhov I~S, Vergeles S~S and Turitsyn S~K 2014 {\it Phys. Rev. Lett.} {\bf 113} 230602

%==================7
\bibitem{Reznic:2017IEEE}
Terekhov I~S,  Reznichenko A~V 2018 {\it IEEE International
Symposium on Information Theory. Proceedings Volume} \textbf{2018}
186

%==================8
\bibitem{TTKhR:2015}
Terekhov I~S,  Reznichenko A~V,  Kharkov Ya~A and  Turitsyn S~K  2017 {\it Phys. Rev. }  E \textbf{95} 062133

%=================9
\bibitem{Terekhov:2017}
Panarin A~A, Reznichenko A~V and~ Terekhov I S 2016 {\it Phys. Rev.}  E {\bf 95} 012127

%=================10
\bibitem{Terekhov:2018}
Terekhov I~S, Chernykh A~I, Smirnov S.~V and Reznichenko A~V 2018
The $\log\log$ growth of channel capacity for nondispersive
nonlinear optical fiber channel in intermediate power range.
Extension of the model. ({\it Preprint } cs.IT/1810.00513)

\end{thebibliography}
\end{document}